\documentstyle[aps]{revtex}
\begin{document}
\title
{Pseudospin SU(2) Symmetry Breaking,
Charge Density Wave \\
and Superconductivity in the Hubbard Model}

\author
{Shun-Qing Shen and X. C. Xie}

\address
 {Department of Physics, Oklahoma State University, Stillwater, OK 74078}
\maketitle
\begin{abstract}
In this paper, we discuss physical consequences of pseudospin SU(2)
symmetry breaking in the negative-U Hubbard model
at half-filling.
If pseudospin symmetry is spontaneously broken while its
unique subgroup U(1) remains invariant,
it will lead to the charge density wave (CDW)
ground state. Furthermore, if
the U(1) symmetry is also broken, the ground state will
have the off-diagonal long range order (ODLRO), signaling a superconductor.
In this case, CDW and superconductivity coexist to form a supersolid.
Finally,
we show that CDW suppresses, but does not destroy superconductivity.

\noindent
PACS No: 74.65+n, 74.20-z
\end{abstract}

The spontaneous symmetry breaking in the condensed matter systems may
produce some physically observable consequences\cite{Anderson83}.
For example, superfluid or
superconductivity comes from the U(1) symmetry breaking, and ferromagnet
and antiferromagnet  in a
spin isotropic system comes from the spontaneous spin SU(2) symmetry
breaking.
Pseudospin SU(2) symmetry firstly discovered in Kondo
lattice model \cite{Jones88} and one-band Hubbard
model \cite{Yang89,Yang90,Zhang90}
is a novel kind of rotation invariance
in the particle-hole space, and has a dual relationship with the usual
spin symmetry \cite{Shen93,Pu94}. It has intrigued
a lot of interests among theoretical
physicists and many studies have been carried out in this
direction.
By using pseudospin operators, Yang\cite{Yang89} has
proposed the so-called ``$\eta$-pairing''
mechanism of superconductivity and his
idea was later realized in the ground
states of the negative-U case
\cite{Shen93,Singh91}.
Zhang has discussed the physical consequences of pseudospin SU(2)
symmetry breaking in the
``$\eta$-pairing'' superconductor \cite{Zhang90}.
Another important outcome of pseudospin SU(2) symmetry is the coexistence
of the charge density
wave and superconductivity, namely supersolid.
By using the partial particle-hole transformation \cite{Shiba72},
it is shown that antiferromagnetic ground state with positive U at
half filling corresponds to a supersolid of negative U. Therefore,
in the negative Hubbard model there is a possible
coexistence of superconductor characterized by the ODLRO
and charge density wave (CDW) characterized by the diagonal long-range order
(DLRO).
Usually, ODLRO is a characteristic of superconductivity,
but in a supersolid with both ODLRO and DLRO whether this is still
true is an open question.
Authors in Ref.\cite{Soni93} have concluded
that the critical magnetic field for
a supersolid with pseudospin symmetry
must be zero since superconductor can freely rotate to
the CDW state because of the SU(2) symmetry and degeneracy.
This leads to the question of understanding the superconducting properties in
a supersolid.

In this paper, we discuss some physical consequences of
pseudospin SU(2)
symmetry in the negative-U Hubbard model.
First, we show that
if pseudospin symmetry is spontaneously broken while its
unique subgroup U(1) remains invariant, it will lead to the CDW
ground state, and the appearance of two massless collective modes according
to Goldstone theorem \cite{Goldstone60}. Second, if
the U(1) symmetry is also broken, the ground state will
have the ODLRO, signaling a superconductor.
In this case, the ODLRO and CDW coexist and form a supersolid.
U(1) symmetry breaking will also give rise to a third branch of massless mode.
Finally, we discuss the superconducting properties in
a supersolid.
We show that CDW suppresses, but does not destroy superconductivity, and
supersolid exhibits
both Meissner
effect and zero resistance.
We also calculate the penetration depth in terms of
the order parameters for superconductor and CDW.
It should be emphasized that the traditional view
considering CDW as produced by breaking of the translational symmetry
which is a discrete symmetry
in the Hubbard model. The significance of our work is
that we find additional continuous symmetry for CDW which has important
physical consequences (like Goldstone bosons).
Our finding of the relation between CDW and pseudospin may not restrict
to the Hubbard model. For example, the pseudospin symmetry is also
exist in the Su-Schrieffer-Heeger hamiltonian\cite{Heeger88}
for describing one-dimensional
polymer and CDW in that system can also be considered
as breaking of pseudospin symmetry.

The one-band Hubbard model on a bipartite lattice $\wedge$ of $A$ and $B$
is defined by
the Hamiltonian
\begin{equation}
H=-t\sum_{\langle ij\rangle,\sigma}c^{\dagger}_{i\sigma}c_{j\sigma} +
\frac{2}{3}U\sum_{i\in \wedge}\tilde{{\bf S}}_i\cdot\tilde{{\bf S}}_i
+2\mu\sum_{i\in \wedge}\tilde{{\bf S}}_i^z
\end{equation}
where $c^{\dagger}_{i\sigma}$ and $c_{i\sigma}$ are the creation and
annihilation operators of electron with spin $\sigma$ at site $i$,
respectively.
The summation of $\langle ij\rangle$ runs over all possible nearest neighbour
pairs. $\mu$ is the chemical
potential and determined by the electron density.
$\tilde{{\bf S}}_i$ is pseudospin
operator and is defined by
\begin{equation}
\left\{
\begin{array}{l}
\tilde{\bf S}^+_i=\epsilon(i)c_{i\uparrow}c_{i\downarrow} \\
\tilde{\bf S}^-_i=\epsilon(i)c^{\dagger}_{i\downarrow}
c^{\dagger}_{i\uparrow}\\
\tilde{\bf S}^z_i=\frac{1}{2}(1-n_{i\uparrow}-n_{i\downarrow})
\end{array}
\right.
\end{equation}
where $\epsilon(i)=+1$ for $i\in  A$, and -1 for $i\in  B$. They
obey the commutation relation of the spin operators.
The total pseudospin operator is
$\tilde{{\bf S}}=\sum_i\tilde{{\bf S}}_i$, and its
three components can be used as the generators to construct a SU(2) group.
The Hamiltonian in Eq.(1) has the commutation relation,
\begin{equation}
[H,\tilde{\bf S}^{\pm}]=\mp 2\mu\tilde{\bf S}^{\pm};\ \
[H,\tilde{\bf S}^z]=0
\end{equation}
Therefore, when $\mu=0$, H has pseudospin SU(2) symmetry, and when $\mu\neq 0$,
the symmetry is explicitly broken by the term containing the chemical
potential. The
latter is very similar to the isotropic spin Heisenberg model in the presence
 of an external magnetic field. On the other hand,
$\tilde{{\bf S}}^z$ always commutes
with the Hamiltonian. Thus, the total number of the electrons is conserved, and
the Hamiltonian always possesses U(1) symmetry, which is the unique subgroup of
SU(2).
In this paper, we are interested in the physical consequences of spontaneous
SU(2)
symmetry breaking and for our purpose we shall only discuss the case with
$\mu=0$. In this case, if the ground state does not possesses
the SU(2) symmetry as the Hamiltonian has, the spontaneous
symmetry breaking occurs.

{}From the structure of SU(2) group, there are two possibilities
for the pseudospin symmetry being spontaneously broken:
i) the SU(2) symmetry is breaking
while its
subgroup U(1) remains invariant; and ii) U(1) symmetry is also
broken. We show that Case i) will produce the CDW ground state.
{}From the definition of pseudospin operator, its z-component is
related to the charge
density, and its corresponding long-range order
is CDW. The CDW state at half
filling with the order
parameter
$\frac{1}{\sqrt{N_{\wedge}}}
\tilde{S}^z_{Q}=\frac{1}{N_{\wedge}}\sum_{i}\epsilon(i)
\langle\tilde{{\bf S}}^z_i\rangle
=-\frac{1}{2N_{\wedge}}\sum_{{\bf k},\sigma}
\langle c^{\dagger}_{{\bf k},\sigma}c_{{\bf k}+{\bf \pi},\sigma}
+c^{\dagger}_{{\bf k}+\pi,\sigma}c_{{\bf k},\sigma}
\rangle$ can be expressed as
\begin{equation}
|CDW\rangle = \prod_{{\bf k},\sigma} (u_k  c^{\dagger}_{{\bf k},\sigma}
+v_k c^{\dagger}_{{\bf k}+{\bf \pi}, \sigma})|0\rangle\
\end{equation}
with $\frac{1}{N_{\wedge}}\sum_{{\bf k}}u_{{\bf k}}v_{{\bf k}}
\neq 0$,
where the {\bf k} is within half of the Brillouin zone.
Although we have
$\langle CDW\vert \tilde{{\bf S}}^{\pm}\vert CDW\rangle
=\langle CDW\vert\tilde{{\bf S}}^z\vert CDW\rangle=0$,
but only $\tilde{{\bf S}}^z\vert CDW\rangle= 0$ and
$\tilde{{\bf S}}^{\pm}\vert CDW\rangle\neq 0$. The SU(2)
symmetry in a state has
been broken unless all eigenvalues for the operator $\tilde{{\bf S}}$ are
zero.
Therefore, $\vert CDW\rangle$ keeps the conservation of the
particle number (or U(1) symmetry)
and destroys the SU(2) symmetry at the same time.

One of the strong evidences to support CDW as product of pseudospin
symmetry breaking is the existence of massless modes, i.e., Goldstone bosons.
According to Goldstone theorem,
when a continuous symmetry is spontaneously broken,
there must exist some massless
modes. The appearance of CDW will produce two massless modes.
To see this, we first come to consider the following correlation functions
\begin{eqnarray}
F_+(t,t')&=&-i\theta(t-t')\langle[\tilde{{\bf S}}^{+}(t), \sum_i \epsilon(i)
\tilde{{{\bf S}}}^-_i(t')]\rangle, \\
F_-(t,t')&=&-i\theta(t-t')\langle[\tilde{{\bf S}}^{-}(t), \sum_i \epsilon(i)
\tilde{{\bf S}}^+_i(t')]\rangle\ ,
\end{eqnarray}
where $\langle \cdots \rangle$ implies the thermodynamic average.
By using the commutation relation between
pseudospin operators and the Hamiltonian, we have
\begin{equation}
F_{\pm}(\omega)=\frac{2\langle\sum_i \epsilon(i) \tilde{{\bf S}}^z_i\rangle}
{\omega\mp 2\mu +i\delta}
\end{equation}
If the CDW order parameter
$\langle\sum_i \epsilon(i) \tilde{{\bf S}}^z_i
\rangle/N_{\wedge}
\neq 0$ ,
then there are two collective modes $\omega_0=\pm 2\vert\mu\vert$. It
should be noticed
that if the chemical potential $\mu$ is not equal to zero, i.e., the symmetry
is
explicitly
broken, the two collective modes are massive.
If the CDW state arises at half filling case which
are our interests in this paper, then $\mu=0$ and
there appear two massless modes according to Eq.(7).
Eq.(7) illustrates that it is the symmetry breaking CDW ground state
which produces
the two massless modes.
We should point out that only one branch is directly observable because
the other one is of negative energy.

The Case (ii) will lead to the coexistence of CDW and superconductivity.
As is well known, the U(1) symmetry breaking
gives rise to a superconductor.
The BCS superconducting state with an order parameter
$\langle c^{\dagger}_{{\bf k},\uparrow}
c^{\dagger}_{-{\bf k},\downarrow}\rangle$ can be written as
\begin{equation}
\vert BCS \rangle=\prod_{{\bf k}}(u_{{\bf k}}+v_{{\bf k}}c^{\dagger}_{{\bf k},
\uparrow}c^{\dagger}_{-{\bf k},\downarrow})\vert 0\rangle\ ,
\end{equation}
here the {\bf k} is within  Brillouin zone.
Clearly, $\vert BCS \rangle$
breaks the conservation of the particle number since $\vert BCS \rangle$
is not an eigenstate of $\tilde{{\bf S}}^z$.
The superconductivity in a system with pseudospin symmetry
can be understood as following.
Pseudospin operators, like the usual spin-1/2 operators, can be regarded as
operator for hard-core bosons. The raising and lowing operators consist
of the local
singlet pairs of electrons. Obviously, no two pairs can occupy one site due to
the Pauli principle. If the hard core bosons condensate macroscopically
at a low temperature,
it gives rise to superconductivity ( the boson here has charge
$2e$). If the condensation happens at the momentum ${\bf q}={\bf \pi}$, the
order parameter can be written as $\langle\frac{1}{N_{\wedge}}\sum_i\exp{(i{\bf
q}\cdot{\bf \pi})}\tilde{{\bf S}}_i^+ \rangle=
\frac{1}{N_{\wedge}}\sum_{{\bf k}}\langle
c_{{\bf k}\uparrow}c_{-{\bf k}\downarrow}\rangle$, which is that of the usual
BCS superconductor. If the condensation happens at ${\bf q}=0$, its
order parameter is $\langle \frac{1}{N_{\wedge}}\sum_i
\tilde{{\bf S}}^+_i\rangle=
\frac{1}{N_{\wedge}}
\sum_{{\bf k}}\langle c_{{\bf k}+{\bf \pi}\uparrow}c_{-{\bf k}\downarrow}
\rangle$. In this case, pseudospin obeys the relationship $\tilde{S}^2-
\tilde{S}^2_z=O(N_e^2)$, which is the criteria for the $\eta$-pairing
superconductor as Yang proposed \cite{Yang89}. Such a phenomenon only happens
in the ground state of the negative U case with $2N_A>N_e>2N_B$, and there,
it is found that $|\langle \sum_i \epsilon(i)\tilde{{\bf S}}_i\rangle|>
|\langle\sum_i\tilde{{\bf S}}_i\rangle|$.
Thus, the two kinds of ODLROs must coexist in the
$\eta$-pairing superconductor. When the condensation occurs, adding or
decreasing a boson in the condensate does not change the total energy. Thus
the chemical potential $\mu$ is equal to zero. Furthermore, the condensation
also breaks the U(1) symmetry in the thermodynamic limit. Therefore, {\it
the local
pairing superconductivity is product of
both pseudospin SU(2) and U(1) symmetries
spontaneously breaking}.
The U(1) breaking should give rise to  another massless
mode \cite{Schrieffer64}.
Together with the two massless modes produced by CDW, there are three
massless modes while U(1) symmetry is also broken.
Since the U(1) group is the unique subgroup of SU(2), the SU(2) symmetry
must be also broken when U(1) is broken. Therefore, in this case, the CDW and
superconductivity must coexist. {\it The SU(2) and U(1) symmetry
spontaneously breaking
gives rise to a supersolid}.
Furthermore, the $T_c$ for CDW is always higher than the $T_c$ for
superconductivity since U(1) breaking always happens after SU(2) symmetry
breaking.

Next we come to discuss the the diamagnetism and resistance in a supersolid
as a product of spontaneous pseudospin symmetry breaking
based on the mean-field theory.
CDW has the diagonal long-range order
and it might suppress the superconductivity
characterized by ODLRO. However, whether it could destroy
superconductivity completely
is an interesting and open question.
In the mean field theory, the Hamiltonian for the
coexistence of CDW and superconductivity
is reduced to
\begin{equation}
H=\sum_{{\bf k},\sigma}\gamma_{{\bf k}}c_{{\bf k},\sigma}^{\dagger}c_{{\bf
k},\sigma}
+\sum_{{\bf k}}(\Delta^*c_{-{\bf k},\downarrow}c_{{\bf k},\uparrow}
+\Delta c_{{\bf k},\uparrow}^{\dagger}c_{-{\bf k},\downarrow}^{\dagger})
-\rho\sum_{{\bf k}}c_{{\bf k},\sigma}^{\dagger}c_{{\bf k}+{\bf \pi},\sigma}
+N_{\wedge}\frac{3}{2\vert U\vert}(\Delta^2+\rho^2),
\end{equation}
where $\Delta=\frac{2}{3}\frac{1}{N_{\wedge}}\frac{1}{U}\sum_ie^{i{\bf
\pi}\cdot
 {\bf r}_i}
\langle \tilde{{\bf S}}_i^+\rangle$ and
$\rho=\frac{2}{3}\frac{1}{N_{\wedge}}\frac{1}{U}\sum_ie^{i{\bf \pi}\cdot {\bf
r}
_i}
\langle \tilde{{\bf S}}_i^z\rangle$. The kinetic energy is $\gamma_{{\bf k}}=
-2t(\cos k_x +\cos k_y +\cos k_z)$. In this approaximation, one
finds that the ground state energy, kinetic energy and equation of order
parameters depend only on the parameter
$R=\sqrt{\Delta^2+\rho^2}$, and
the states
with pure CDW and pure superconductivity
are degenerated. However, we should point out that
the supersolid has a unique non-degenerate ground state as shown
beyond the mean-field theory \cite{Pu94,Lieb89}.

Generally speaking, Meissner effect and zero resistance are
two hallmarks of a superconductor.
In this model, the current operator is
\begin{equation}
{\bf j}_0 = iet\sum_{i,\vec{\delta},\sigma}\vec{\delta}c^{\dagger}_{i+{\bf
\delta},\sigma}
c_{i,\sigma},
\end{equation}
where $\vec{\delta}$ is the lattice vector for the nearest neighbour sites.
When the system is in the presence of a weak magnetic field,
the linear response theory \cite{Schrieffer64}
tells us that the response current can be expressed by
\begin{equation}
{\bf j}={\bf j^d} +{\bf j^p},
\end{equation}
where
\begin{equation}
j^d_{\mu} = -te^2\sum_{i,\sigma}\langle g\vert c^{\dagger}_{i+{\bf
e}_{\mu},\sigma}
c_{i,\sigma}+c^{\dagger}_{i,\sigma}c_{i+{\bf e}_{\mu},\sigma}\vert g\rangle
A_{\mu}({\bf r}_i,t),
\end{equation}
and
\begin{equation}
j^p_{\mu} = +i\sum_{i,i',\nu}\int dt'\langle g\vert[j_{0,\mu}({\bf r}_i,t),
j_{0,\nu}({\bf r}_i',t')]\vert g\rangle \theta(t-t')A_{\nu}({\bf r}_i',t').
\end{equation}
${\bf e}_{\mu}$ is a unit vecor along the direction $\mu=(x,y,z)$
and $\vert g\rangle$ is the ground state in the absence of the field.
Furthermore,
\begin{equation}
j_i({\bf q}, \omega)=-e^2[\langle -K_i\rangle - R_{ij}({\bf q},
\omega)]A_{j}({\bf q}, \omega)
\end{equation}
where
\begin{equation}
\langle K_i\rangle = -\frac{2}{3}\frac{1}{N_{\wedge}}\sum_{{\bf
k}}\frac{\gamma^2_{{\bf k}}}{\omega_{{\bf k}}}
\end{equation}
and
\begin{equation}
R_{ij}({\bf k},\omega)=-i\int_{-\infty}^{\infty}\frac{dt}{2\pi}
e^{i\omega t}
\langle g\vert [j_i({\bf q},t),j_j(-{\bf q}, 0)]
\vert g\rangle \theta(t).
\end{equation}
The superfluid density is determined by \cite{Schrieffer64,Scalapino93}
\begin{eqnarray}
\left ( \frac{n_s}{m}\right)^*&\equiv&\langle -K_i\rangle -
R_{ii}(q_i=0,q_{j\neq i} \rightarrow 0,
\omega=0)\nonumber \\
&=&8t^2\frac{1}{N_{\wedge}}\sum_{{\bf k}} \sin^2 k_i\frac{\Delta^2}{\omega^3}
+\left ( \frac{n_s}{m}\right)_{CDW}
\end{eqnarray}
where $\left ( \frac{n_s}{m}\right)_{CDW}$ is the superfluid density
for a pure CDW state with
$\rho=R$ and $\Delta=0$. The pure CDW state here is an insulator
\cite{Scalapino93} with
\begin{equation}
\left ( \frac{n_s}{m}\right)_{CDW}=
\frac{2}{3}\frac{1}{N_{\wedge}}\sum_{{\bf k},\sigma}
\frac{\gamma^2_{{\bf k}}}{
\omega_{{\bf k}}} -8t^2\frac{1}{N_{\wedge}}\sum_{{\bf k}}\sin^2k_x\frac{R^2}
{\omega^3_{{\bf
k}}}=0.
\end{equation}
Therefore, the superfluid density
 $\left ( \frac{n_s}{m}\right)^*  $ for the supersolid
ground state with both CDW and superconductivity
is proportional to $\Delta^2$ and is always positive and nonzero
as long as $\Delta\neq0$.
Furthermore, as the quasiparticle spectrum
has a finite energy gap $R$,
the Drude weight is equal to the superfluid weight \cite{Scalapino93}
\begin{equation}
 \frac{D}{\pi e^2}\equiv\left ( \frac{n}{m}\right)^*
\equiv\langle -K_x\rangle-Re\{R_{ii}(q=0,
\omega\rightarrow 0)\}=\left ( \frac{n_s}{m} \right )^* \neq 0
\end{equation}
which implies the zero resistance.
The penetration
depth for a supersolid is $\lambda_s=(4\pi n_s e^2/mc^2)^{-1/2}$  and
$\lambda_c=(\frac{4\pi e^2}{3c^2 N_{\wedge}}\sum_{{\bf k}}
\frac{\gamma_{{\bf k}}^2}{\omega_{{\bf k}}})^{1/2}$ is
the penetration depth for a pure superconductor
with $\rho=0$ and $\Delta=R$. We have
\begin{equation}
\lambda_s=(1+\frac{\rho^2}{\Delta^2})^{1/2}\lambda_c\geq\lambda_c.
\end{equation}
This expression indicates that the perfect diamagnetism arises as soon as ODLRO
is present,
i.e., $\Delta\neq 0$. When $\Delta\rightarrow 0$,  $\lambda_s$
approaches infinite,
and the diamagnetism disappears. Hence,
the order parameter $\Delta$ guarantees the superconductivity
with both Meissner effect and zero resistance. $\lambda_s$
 will increase with the CDW order parameter
$\rho$, thus, in the supersolid
the CDW suppresses the superconductivity, but does not
destroy it completely.
In the negative U case on a cubic lattice,
the SU(2) symmetry
ensures that the ratio for $\rho$ and $\Delta$
is fixed at $\rho/\Delta=1/\sqrt{2}$ \cite{Pu94},
which gives the value
$\lambda_s=\sqrt{\frac{3}{2}}\lambda_c$. Therefore,
we conclude that the supersolid is a true superconductor, not an
insulator as being claimed in the previous studies Ref.\cite{Soni93,Tian94}.

Before ending the paper, we would like to discuss some experimental
implications of our results.
The following discussions are only meaningful if
the experimental samples can be approximately described
by hamiltonians which
contained the pseudospin SU(2) symmetry.
The coexistence of CDW and superconductivity has been observed in
several materials \cite{Smith74}.
One common feature is that the critical temperature for the CDW
is always greater than that for
superconductivity. This phenomenon can be understood very naturally
in our theory.
Superconductivity is the product of U(1) symmetry breaking,
and the SU(2) symmetry
should be broken (producing CDW) before its subgroup U(1) is broken.
Therefore, the critical
temperature for CDW
must not be less than that for superconductivity. The CDW
has DLRO
and always suppresses ODLRO. But the SU(2) symmetry guarantees
the coexistence of both
long-range orders at low temperatures.

The collective modes are the products of the spontaneous symmetry
breaking. There are two branches of massless modes
accompanying the appearance of CDW,
but only one is  experimentally observable
due to the negative
energy in the other branch for a finite momentum.
After U(1) symmetry breaking,
a new branch of massless mode should appear.
The massless mode and massive mode correspond to the phase mode and
amplitude mode in order parameter, thus, usually appear in pair \cite{Lee74}.
The massive modes have been observed experimentally \cite{Sooryakumar80}.
The number of modes is
one if the system is CDW ground state and two if CDW is coexistent
with superconductor, these numbers agree with our findings.

In summary, the CDW in the Hubbard model
is the product of spontaneous pseudospin SU(2) symmetry
breaking, and furthermore if its subgroup U(1) symmetry is also broken, there
is
coexistence of both
CDW and superconductivity.
The additional global and
continuos symmetry breaking presents important and physically observable
consequences, and makes a natural understanding of a supersolid.

\end{document}